 \pdfoutput=1
\documentclass{article}
\usepackage[pdftex]{graphicx}
\usepackage{epstopdf}
 \usepackage{ifpdf}
\usepackage[numbers]{natbib}
\usepackage{algorithm}
\usepackage{arevmath}     
\usepackage[noend]{algpseudocode}
\usepackage{placeins}

\usepackage{fixltx2e}
\usepackage[a4paper,textwidth=18cm,textheight=24cm,top=2.85cm, bottom=2.85cm, left=1.5cm, right=1.5cm]{geometry}
\usepackage{icdp2009}
\graphicspath{{examples/images/}}
\makeatletter
\long\def\@makecaption#1#2{%
\vskip\abovecaptionskip
\sbox\@tempboxa{#1. #2}%
\ifdim \wd\@tempboxa >\hsize
#1. #2\par
\else
\global \@minipagefalse
\hb@xt@\hsize{\box\@tempboxa\hfil}%
\fi
\vskip\belowcaptionskip}
\makeatother

\usepackage[none]{hyphenat}
\usepackage[labelformat=empty]{caption}
\usepackage{newtxmath}
\usepackage{times}
\usepackage{indentfirst}
\usepackage{diagbox}
\usepackage{capt-of}
\usepackage{tabu}
\usepackage[justification=centering]{caption}
\usepackage{multicol}
\usepackage{multirow}
\usepackage{array}
    \newcolumntype{P}[1]{>{\centering\arraybackslash}p{#1}}
    \newcolumntype{M}[1]{>{\centering\arraybackslash}m{#1}}
\newcolumntype{L}{>{\centering\arraybackslash}m{3cm}}

\usepackage{booktabs}
\usepackage[dvipsnames]{xcolor}
\usepackage{tikz}
\ifpdf

      \usepackage[pdftex]{hyperref}

    \else

      \newcommand{\href}[2]{#2}

    \fi

\begin{document}
\noindent

\bibliographystyle{ieeetr}
\title{Fault Localization in Cloud using Centrality Measures}
\authorname{\large{Narayanaa S R, Sivaranjan M and  Lekshmi R S}}
\authoraddr{{Department of Applied Mathematics and Computational Sciences} \\ \\  PSG College of Technology, Coimbatore}

\maketitle

\begin{abstract}
 Fault localization is an imperative method in fault tolerance in a distributed environment that designs a blueprint for continuing the ongoing process even when one or many modules are non-functional. Visualizing a distributed environment as a graph, whose nodes represent faults (fault graph), allows us to introduce probabilistic weights to both edges and nodes that cause the faults. With multiple modules like databases, run-time cloud, etc. making up a distributed environment and extensively, a cloud environment, we aim to address the problem of optimally and accurately performing fault localization in a distributed environment by modifying the Graph optimization approach to localization and centrality, specific to fault graphs.
\end{abstract}


\section{INTRODUCTION} 
Cloud computing is the on-demand availability and delivery of various computer system resources, cloud storage and computing power, without requiring direct management by the user. Today, large clouds, predominantly have functions distributed over various locations from central servers. Cloud computing delivers three types of computing resources - hardware, software, and software development framework, also known as a platform. Recently, the extensive use of cloud services for hosting enterprise applications is leading to service reliability and availability issues for both service users and providers. Several types of faults may occur in the cloud environment leading to failures and performance degradation \cite{study}.

Failures, generally lead to breakdown or shut down of a system. In distributed computing architecture, the notion of failure is characterized by partial failures. A fault may occur in any constituent node, process, or network component of the distributed system. This leads to a partial failure and consequently, the performance of the system degrades until there is a complete breakdown. Fault tolerance in a system is the capability of a system to continue performing its expected function regardless of faults. This way, a fault-tolerant system is expected to be reliable and operate successfully in the case of a failure or breakdown. In case of a fault occurring from any particular node, a fault-tolerant system must be able to identify potentially vulnerable modules in a distributed system to replicate their functionality to prevent complete failure of the system.  

State-of-the-art fault localization approaches rely on long training sessions based on simulated faults, which is an expensive practice, since it is hard to inject multiple classes of faults in several machines with different access controls, and iterate this process while the system evolves \cite{survey}. Fault detection using deep learning is also studied in the case of small amount of cloud fault feature data \cite{deep}.

Centrality measures in a graph are a direct indication of influential nodes and links in the network. In most cases, they show the relative significance of nodes and edges in the system. It is a direct indication of influential nodes and links in the network. Researchers have developed many algorithms for identifying relatively important nodes in a network. The basic and locally computed centrality measure is degree centrality. Other commonly used metrics are betweenness centrality, closeness centrality, PageRank, and HITS eigenvector centrality, which are more relevant in single-layer networks, limiting their applications in multi-layer networking systems \cite{centrality}. Among these, PageRank is a popular and frequently used metric, designed for ranking of web pages, it is widely applied in other applications also. 

In this paper, we propose a fault localization method that uses centrality measures and works within the constraints of Cloud systems to help us accurately identify important components in the cloud and localize them in case of a breakdown or failure, triggered by itself or another component it depends on. 

The contributions of this work are: (i) a novel approach to identify highly important and vulnerable components in cloud environments to localize the fault triggering components (ii) an experimental evaluation of the suggested method for identifying and locating different combinations of faults arising from different modules.

The paper is organized as follows. Section 2 overviews the existing work on fault localization techniques. Section 3 discusses the theoretical background required. Section 4 proposes a centrality measure-based localization method. Section 5 illustrates the experimental results of the approach. Section 6 summarizes the main contribution of the paper and a few future work possible.

\section{RELATED WORK}
\indent A failure occurs when the user perceives that the software has ceased to deliver the expected result. The user may need to identify the severity of the levels of failures such as catastrophic, critical, major, or minor, depending on their impact on the systems. Omission, Hardware, Software, Network, Response, Virtual Machine (VM), Application failures are various types of failures in Cloud systems. An abnormal condition or defect at the equipment, component or sub-system level which may lead to a failure is termed as a Fault. Omission faults, Aging-related faults like denial of services, Response faults, Software Faults, Timing faults, Transient faults are various types of faults \cite{faultsd}. The fault localization process is to be undertaken by developers to identify the fault responsible for the failure.
To handle the fault that may occur there are different types of techniques of fault tolerance in the literature that is classified into two categories:

\subsection{Proactive Techniques}

In Proactive Fault Tolerance, the problems are anticipated before they occur and are avoided based on the influence on the task. One of the methods in Proactive Fault Tolerance is preemptive migration, which uses a control system where the tasks are proctored. When a task is found to be vulnerable, it is shifted to another virtual machine instance. The disadvantage of this method is the wait time for calling another virtual machine \cite{pro}.
Another example of a proactive fault-tolerance method is Proactive Co-ordinated fault tolerance or PCFT. The PCFT approach aims at monitoring a deteriorating physical machine. Given a deteriorating physical machine, the PCFT approach searches for some optimal target physical machines for the virtual machines hosted on the deteriorating physical machine \cite{reliable}.

\subsection{Reactive Techniques}

Reactive fault tolerance approaches handle the faults after their occurrence. The working of reactive approaches is response-based rather than anticipation. Reactive approaches are usually conservative and need not examine the system behavior. Steps involved in reactive fault tolerance include (a) Replication: this is a technique used to replicate the task on more than one resource, (b) Task re-submission: when the task fails it will run again in either the same node or on a different resource, this increases the time taken for execution over the expected time because of the repetition of task (c) Retry: re-execution of the failed task in the same resource, but it consumes the same execution time as the Task re-submission, (d) Checkpointing or Restart: It takes snapshots at different times during task execution. This allows the task to be re-executed from the last fault-free state \cite{rea}.
\vspace{2mm}

Other methods including the meta-heuristic-based method have also been used for improving fault tolerance methods in cloud systems to enhance their performance and to make the systems robust \cite{metasd}. Failure risk based Cloud Classification System reliability and efficiency improvement methods have been developed by combining risk identification with proactive actions \cite{ccs}.

\section{THEORETICAL BACKGROUND}
In this section, we discuss some basic concepts about the types of faults in cloud architecture and centrality measures in a graph.

\subsection{Cloud Computing Infrastructure}
Cloud computing infrastructure includes the computers, network facilities, storage devices, and other connected components necessary for offering cloud computing resources and services to users as mentioned in Table 1. These hardware components are located within enterprise data centers. These encompass solid-state drives, multi-core servers, and hard disk drives offering stable storage and network devices, such as firewalls, switches, and routers; all on a large scale \cite{scidir}\cite{netnewfault}.
\vspace{2mm} 
\begin{table}[ht!]
\begin{tabular}{|p{2cm}||p{5.7cm}|}
\hline
\multicolumn{2}{|c|}{Various components in a distributed system} \\
\hline
Servers   & The physical machines that act as host machines for one or more virtual machines. \\
\hline
Virtualization &   Technology that abstracts physical components such as servers, storage, and networking and provides these as logical resources.\\
\hline
Storage & In the form of Storage Area Networks (SAN), network-attached storage (NAS), disk drives etc. along with facilities such as archiving and backup.\\
\hline
Network   & To provide interconnections between physical servers and storage. \\
\hline
Management &  Various software for configuring, management and monitoring of cloud infrastructure including servers, network, and storage devices.\\
\hline
Security& Components that provide integrity, availability, and confidentiality of data and security of information, in general.\\
\hline
\end{tabular}
\caption{Table 1: Various components in a distributed system}
\end{table}
\vspace{2.5mm} 

Many factors could trigger a fault including data corruption, hardware or software or network outage, unreliable clocks,	method failure, parity error, head crash,	denial of service, disk full, etc. Faults, in general, are classified into the below-mentioned categories: 

\subparagraph{Network faults:}
Since cloud computing resources are accessed over the internet, one of the most important causes of failure in cloud computing is network faults. These faults may occur due to partitions in the network, packet loss or corruption, congestion, failure of the destination node or link, etc.

\subparagraph{Physical faults:} Physical faults mainly occur in hardware resources, such as faults in CPUs, in memory, in storage, failure of power etc.

\subparagraph{Process faults:} These faults may occur in processes because of resource shortage, bugs in software, incompetent processing capabilities, etc.

\subparagraph{Service expiry faults:} If the service time of a resource expires while an application that leased it is using it, it leads to service faults. 

\subparagraph{{Why Perform Fault Localization?}} Fault localization is essential to examine the cloud architecture and help developers improve the reliability and safety of information systems. Some of the examples of fault symptoms include unreachable hosts or networks and in most cases, the origin of the path traced by the defect is important to understand the root cause of a fault.

\subsection{Centrality Measures}
In graph theory and network analysis, indicators of centrality identify the most important vertices and edges within a graph. They are scalar values given to each node or edge in the graph to quantify its importance based on some assumptions.  Centrality concepts were originally developed for social network analysis, and many of the terms used to measure centrality reflect their sociological origin. There are a vast number of centrality measures used by analysts. In our approach, we use closeness, eigenvector, and alpha centrality measures.

\subparagraph{Closeness Centrality:}
In a connected graph, closeness centrality (or closeness) of a node is a measure of centrality in a network, calculated as the reciprocal of the sum of the length of the shortest paths between the node and all other nodes in the graph. Thus, the more central a node is, the closer it is to all other nodes \cite{closeness}.

\subparagraph{Eigenvector Centrality:} 
In graph theory, eigenvector centrality is a measure of the influence of a node in a network where relative scores are assigned to all nodes in the network based on the concept that connections to high-scoring nodes contribute more to the score of the node in question than equal connections to low-scoring nodes. A high eigenvector score denotes that a node is connected to many nodes who themselves have high scores \cite{eigen}.

\subparagraph{Alpha Centrality:} 
In graph theory and social network analysis, alpha centrality is an alias to Katz centrality \cite{katz}. It is a measure of the centrality of nodes within a graph. It is an extension of eigenvector centrality where the degree by which the nodes are impacted by the importance of external sources are taken into account. \\

\section{PROPOSED MODEL}
Any graph-theoretic model relies on a precise establishment of relationships between nodes involved in the graph and require profound understanding of the underlying system behavior \cite{recentadvances}. Fault propagation graphs, usually an input to clustering techniques, are common ways to trace the components affected by the failure of a subset of components in a network of nodes \cite{compare}. We define a directed fault graph FG(V, E) whose V represents different faults associated with components and E represents whether or not will occurrence of a fault would trigger another fault. The proposed centrality measure-based fault localization method assigns the importance of modules in a given cloud architecture with computed rank values of different faults associated with a component, obtained from the fault graph model generated. Computation is based on iterations over fault nodes with weights on edges and nodes calculated with certain probabilistic values. 

We initially consider the list of all the cloud components like Runtime, pod, database, etc., labeled C{\textsubscript{1}}, C{\textsubscript{2}}, C{\textsubscript{3}}, ... , C{\textsubscript{n}} that could potentially become faulty. Further, we derive a list of all faults f\textsubscript{k\textsubscript{i}}, the suffix indicating the i\textit{th} fault arising from component k. With this established, one-to-many mapping between components and faults associated with the components is made as shown in Figure 4.1. In any given cloud environment, components in the environment can fail due to multiple faults. Subsequently, a fault could trigger a subset of other faults as the cloud could be seen as one connected graph.

Every fault f\textsubscript{k\textsubscript{i}}, initially, has an independent probability P{\textsubscript{f\textsubscript{k\textsubscript{i}}}} of arising regardless of being triggered by any other fault. Identification of a particular fault in a network is carried out using various failure detection methods including signature and anomaly-based IDS \cite{acceptancepaper}. These probabilistic weights are assigned to all faults which are mapped to cloud modules. These weights are based on the nature of the module and other external factors, which by default are uniform for all the faults and can be assigned based on the log of fault data.

We introduce an impact factor value ifv(f\textsubscript{k\textsubscript{i}}, f\textsubscript{m\textsubscript{j}}) that tells how much of an impact the fault f\textsubscript{k\textsubscript{i}} has on fault f\textsubscript{m\textsubscript{j}} where the relationship from f\textsubscript{k\textsubscript{i}} to f\textsubscript{m\textsubscript{j}} is one-to-one. These values are specific to architectures and they are calculated based on normalized conditional probability values of a pair of faults and the odds of a fault occurring relative to the occurrence of another fault from the log data of faults. These values represent the initial edge weights of the fault graph FG. As mentioned earlier, the nodes of the directed fault graph FG, have values that represent the initial independent probabilities of occurrence of the corresponding fault. Figure 4.2 shows a schematic graph consisting of two distributed components C{\textsubscript{1}} and  C{\textsubscript{2}}, fault f\textsubscript{1\textsubscript{i}} occurring with independent probability 0.179 and f\textsubscript{2\textsubscript{j}} with probability 0.232 are caused by  C{\textsubscript{1}} and  C{\textsubscript{2}} respectively where ifv(f\textsubscript{2\textsubscript{j}}, f\textsubscript{1\textsubscript{i}}) = 0.34 is the impact factor value the fault node f\textsubscript{2\textsubscript{j}} has on f\textsubscript{1\textsubscript{i}}. The influence value is the normalized conditional probability values of the pair f\textsubscript{1\textsubscript{i}} and f\textsubscript{2\textsubscript{j}}
 
\begin{figure}[htb] 
\begin{center} 
\includegraphics[width=6.5cm]{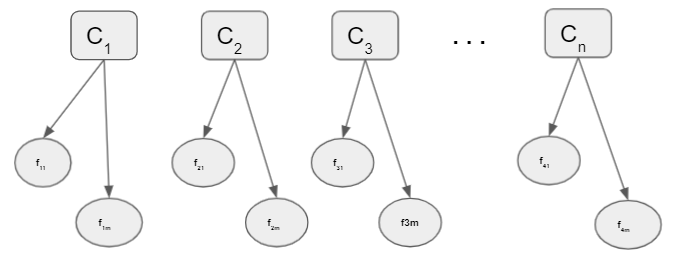}
\end{center} 
\captionsetup{justification=raggedright,singlelinecheck=false}
\caption{\textbf{Figure 4.1 A graphical representation of different modules in a distributed computing model - C\textsubscript{1}, C\textsubscript{2}, C\textsubscript{3}, ..., C\textsubscript{n} where multiple faults - f\textsubscript{11}, f\textsubscript{12}, f\textsubscript{21}, f\textsubscript{22}, etc. are caused by the failure or breakdown of modules C\textsubscript{1} to C\textsubscript{n}} }
\label{fig:curva} 
\end{figure} 

\begin{figure}[H] 
\begin{center} 
  \includegraphics[width=4.5cm]{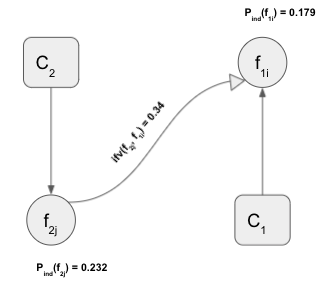}
\end{center} 
\captionsetup{justification=raggedright,singlelinecheck=false}
\caption{\textbf{Figure 4.2 A schematic graph consisting of two distributed components C\textsubscript{1} and C\textsubscript{2} where f\textsubscript{1i} and f\textsubscript{2j} are the faults caused by them respectively where ifv(f\textsubscript{2j}, f\textsubscript{1i}) is the influence fault node f\textsubscript{2j} has on f\textsubscript{1i}}}
\label{fig:curva} 
\end{figure} 

Subsequently, the edge weight of any edge from f\textsubscript{1} to f\textsubscript{2} with an initial edge weight, impact factor {ifv(f\textsubscript{1}, f\textsubscript{2})} and the node weight of a node, P\textsubscript{fi}, representing the probability of occurrence of the fault it represents, f\textsubscript{2}, impacted by f\textsubscript{1} in the fault graph FG are updated to
\[
E\textsubscript{f1, f2} = P\textsubscript{f\textsubscript{1}} * ifv\textsubscript{f\textsubscript{1},f\textsubscript{2}} + Z,
\]
\[
P\textsubscript{f2} = P\textsubscript{f\textsubscript{1}} * ifv\textsubscript{f\textsubscript{1},f\textsubscript{2}} 
\]
where {Z} representing normalization constant which is added to all the nodes that are involved in a cycle to stop the iteration when updating the node weights is \\

\begin{center}
  \includegraphics[width=7cm]{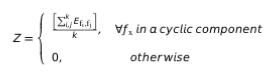}
\end{center}

Different edges in the fault graph have different edge weights which are differentiated using edge thickness as shown in Figure 4.3. It is experimentally seen that the efficiency and the accuracy of the node finding algorithm depend on the accuracy of the probabilistic assumption of each node. 

\begin{figure}[H] 
\begin{center} 
  \includegraphics[width=7cm]{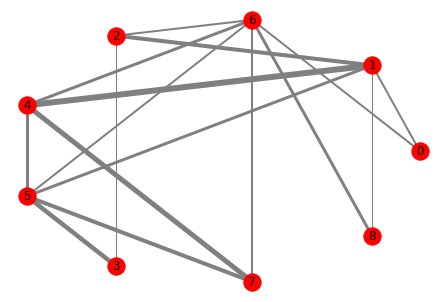}
\end{center} 
\captionsetup{justification=raggedright,singlelinecheck=false}
\caption{\textbf{Figure 4.3  A graphical representation of subset of fault node graph with faults of different components connected with probabilistic edge weight denoted by networkX thickness}} 
\label{fig:curva} 
\end{figure} 

\subsection{Identification of nodes based on centrality measures}
Most site reliability engineering teams use monitoring and alerting systems in which a node's signal value is used to determine the subsequent actions to be taken. The four golden signals in monitoring are traffic, latency, saturation, and errors. In our model, we record the cause and signal values emitted by the components and we set threshold values for the signal values emitted by the cloud components, and whenever the signal values cross the threshold value, with the type of fault recorded, fault propagation graph FP from the original directed fault graph FG, is initialized as shown in Figure 4.4. 

\begin{figure}[H] 
\begin{center}
  \includegraphics[width=5cm]{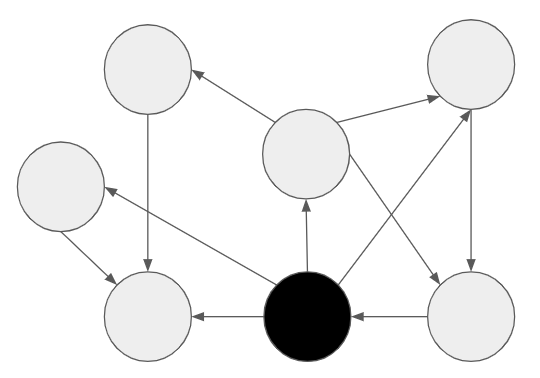}
\end{center}
\captionsetup{justification=raggedright,singlelinecheck=false}
\caption{\textbf{Figure 4.4 A schematic graph in which black colored node represents a potential faulty node from which a fault propagation graph is constructed}}
\label{fig:curva} 
\end{figure} 

We run eigenvector-based and closeness-based centrality indices on the fault propagation graph to find the possibly faulty nodes. With the nodes indicating the type of fault with the possibility of occurrence known, we can then find the vulnerable cloud components from the mapping between faults and cloud modules. For every fault node $f\textsubscript{j}$ $\in$ V, the centrality measure values are calculated using the three centrality measures,

\[ EVR(f\textsubscript{n}) = \sum_{f\textsubscript{j} \in V}^{}
\frac{EVR(f\textsubscript{j})}{L(f\textsubscript{j})} * ifv(f\textsubscript{j}, f\textsubscript{n})\]

where EVR(f\textsubscript{n}) is the eigenvector based rank for a fault node f\textsubscript{n}, and L(f\textsubscript{j}) is the number of nodes that are connected to f\textsubscript{j} and,

\[ CR(f\textsubscript{n}) = \sum_{f\textsubscript{j} \in V}^{}\frac{1}{d(f\textsubscript{n}, f\textsubscript{j})} * ifv( f\textsubscript{n}, f\textsubscript{j}) \]
where CR(f\textsubscript{n}) is the closeness centrality based rank for a fault node f\textsubscript{n}, calculated for any node f\textsubscript{j} that f\textsubscript{n} connects to and d(f\textsubscript{n}, f\textsubscript{j}) is the shortest distance between f\textsubscript{n} and f\textsubscript{j}.

\vspace{2mm} 
Different choices of centrality measure rank fault nodes differently. With the important nodes, representing the faults of high importance and high likelihood of occurring ranked and number of nodes to consider for fault localization known, the faults can be mapped back to the components that raise them to know which get an insight of whose replicas are needed \cite{replica}.\\

Alpha Centrality on weighted fault network graph begins with the identification of high centrality score nodes using alpha centrality measure where we consider the weighted adjacency matrix A{\textsubscript{FP}} of the fault propagation graph generated where the edge weight between any f{\textsubscript{i}} and f{\textsubscript{j}}, W\textsubscript{E\textsubscript{i,j}}, denotes the degree of influence of f{\textsubscript{i}} on f{\textsubscript{j}}.
\begin{figure}[H] 
\begin{center} 
  \includegraphics[width=6.5cm]{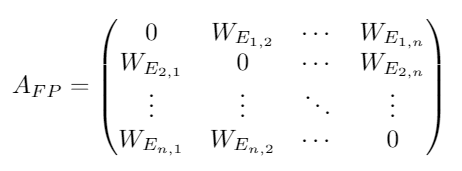}
\end{center} 
\label{fig:curva} 
\end{figure} 
With the weighted adjacency matrix for the fault graph FG known, alpha centrality measure values for the nodes representing faults is calculated. Alpha centrality measure for the graph FG is given by,
\[ C\textsubscript{Katz}(f\textsubscript{i}) = \alpha \sum_{j = 1}^{n} A\textsubscript{FG}({\textsubscript{j, i}}) . C\textsubscript{Katz}(f\textsubscript{j}) + \beta \]
where C\textsubscript{Katz}(f\textsubscript{i}) is the alpha-centrality based rank for a fault node (f\textsubscript{i}), A{\textsubscript{FG}} represents the adjacency matrix of the fault propagation graph, $\alpha$ is called attenuation factor that lies within the range [0, 1] and $\beta$ is the initial weight given to vertex as a way to differentiate vertices using some quality not modelled by adjacencies, which in our case is the probability of occurrence. Rewriting the equation in a vector form, we get, 
\[ 
C\textsubscript{Katz} = \alpha A\textsuperscript{T} C\textsubscript{Katz} + \beta 
\]
\[
C\textsubscript{Katz} = \beta (I - \alpha A\textsuperscript{T})\textsuperscript{-1} 
\]
When $\alpha$ = 0, the eigenvector centrality is nullified and all the nodes have the same centrality value $\beta$. It can also be observed that as $\alpha$ value increases, the impact of $\beta$ reduces. We select 0 $\leq$ $\alpha$ $<$ \( \frac{1}{\lambda} \), where $\lambda$ represents the numerically largest eigenvalue of A\textsuperscript{T}. In the case of alpha centrality, calculation of node centrality values in weighted networks lets us consider the edge weights which is a direct measure of the degree of influence of a fault on another fault.

\section{EXPERIMENTAL RESULTS} 

We conducted tests with over a hundred simulated faults originating from various cloud modules. With the module-fault mapping established, initial probabilities of fault occurrence for independent faults were taken from the recent datasets \cite{loca}. Further, the impact factor mapping between fault nodes was assigned based on the frequency of dependency of one fault triggering module over another. With these values computed and the faulty node from which the fault propagation graph is constructed known, the relative importance of fault nodes are calculated based on the three centrality measures. In a real-time system, this could be input to what requires replication for fault tolerance \cite{replica}. Table 2 represents the list of simulated faults considered along with the module they originate from. The results of the computation directly relate to the modules that are more vulnerable than the rest, in an occasion of a fault occurring. \\

\begin{table}[ht!]
\begin{tabular}{|c|c|}
 \hline
 \multicolumn{2}{|c|}{Fault List} \\
 \hline
 Fault &  Module\\
 \hline
 log.fault.vm.steps.smf.svc.maintenance &  VM\\
 log.fault.vm.steps.cpu.generic-sparc.strand & VM\\
 log.fault.proxyDeployment.steps.InvalidPattern & Proxy\\
 {log.fault.runtime.steps.jwt.KeyParsingFailed} & Runtime\\
 . . . & \\
 log.fault.dbauth.steps.QuotaViolation & Database\\
 log.fault.runtime.steps.jwt.TokenNotYetValid & Runtime\\	
 \hline
\end{tabular}
\caption{Table 2: List of faults and the module they originate from}
\end{table}

\begin{table*}
\renewcommand{\arraystretch}{0.2}
    \centering
    \begin{tabular}{m{2.5cm}m{4.5cm}m{4.5cm}m{4.5cm}}
    
       \toprule
        \diagbox[width=8em]{\\ \\ \\ Centrality}{Threshold} & \hspace{20mm}0.6 & \hspace{20mm} 0.7 & \hspace{20mm} 0.8 \\
        \midrule
        (a) Closeness & 
        \includegraphics[width=45mm]{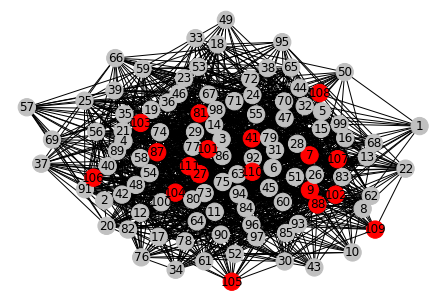}  
        &\includegraphics[width=45mm]{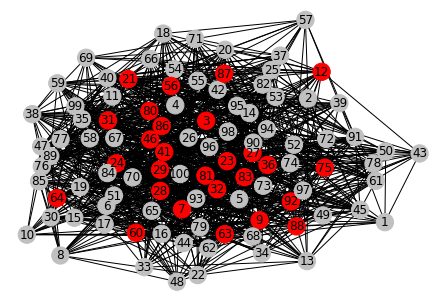}  
        &\includegraphics[width=45mm]{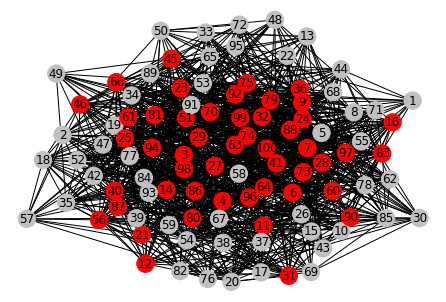}  \\
        
        (b) Eigen & 
        \includegraphics[width=45mm]{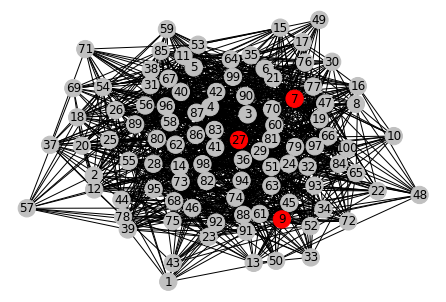}  
        &\includegraphics[width=45mm]{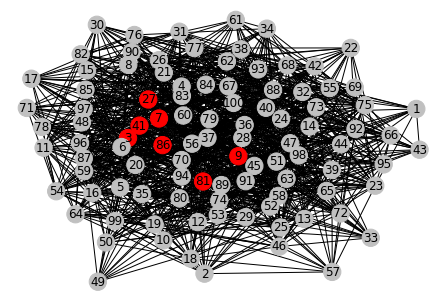}  
        &\includegraphics[width=45mm]{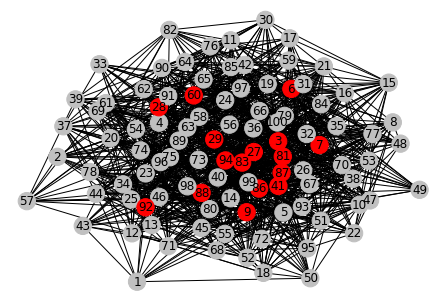} \\
       
        (c) Alpha & 
        \includegraphics[width=45mm]{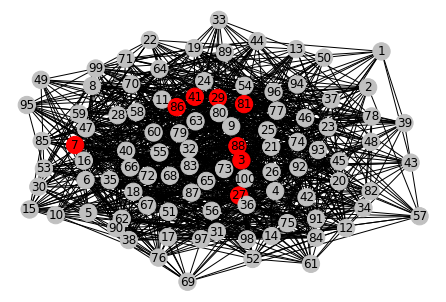}  
        &\includegraphics[width=45mm]{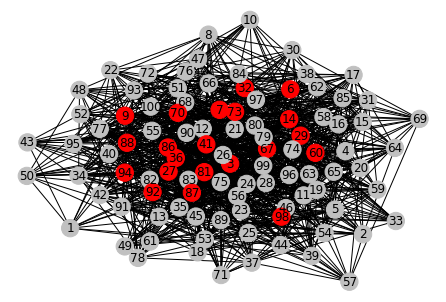}  
        &\includegraphics[width=45mm]{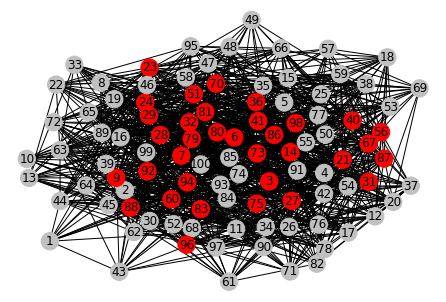} \\
        \bottomrule
    \end{tabular}
    \caption*{\textbf{Figure 5.1 Graphical representation of fault node graphs with 100 faults using networkX}}
    \label{tbl:table_of_figures}
\end{table*}

\begin{table*}
\centering
\begin{tabu}to \textwidth {X[c]X[c]X[c]}

  \includegraphics[width=45mm]{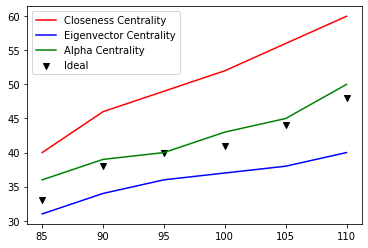}\captionof*{figure}{Threshold = 0.6}    &\includegraphics[width=45mm]{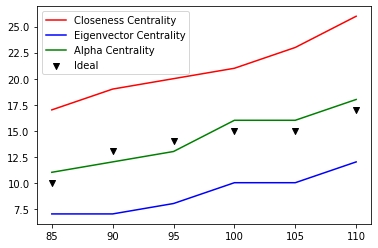}\captionof*{figure}{Threshold = 0.7} &\includegraphics[width=45mm]{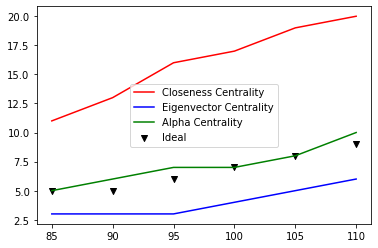}\captionof*{figure}{Threshold = 0.8}\\

\end{tabu}
\caption*{\textbf{Figure 5.2 Graphical comparison of Closeness, Eigenvector and Alpha \\centrality measures for various threshold values}}
\end{table*}

Vulnerable modules were found from the faults that were most likely to occur based on the centrality rankings for different threshold values. The test was run with 85, 90, 95, 100, 105, 110 faults and vulnerable modules were observed for threshold values of 0.6, 0.7, and 0.8 using the three centrality measures. Figure 5.1 shows the graphical representation of the fault node graph constructed using networkX with 100 faults considered for closeness centrality, eigenvector centrality, and alpha centrality for threshold values of 0.8, 0.7, and 0.6 respectively. It is observed that alpha centrality finds the vulnerable modules for all the considered threshold values more accurately than eigenvector and closeness centrality measures. Figure 5.2 shows the comparison of observations with the three different centrality measures for threshold values 0.6, 0.7, and 0.8 respectively, against the ideal values. This high accuracy observed in alpha centrality can be explained by the attribution given to the edge weight calculated with the degree of impact the failure of one node has on the other. It is also observed that the closeness centrality measure produces the highest number of false-positive faults among the three considered centrality measures. Since the efficiency with which the faults propagate is less relevant in finding the vulnerable fault modules, a high number of false-positive cases are observed in closeness centrality measure. The accuracy values recorded for the considered faults using alpha, eigenvector, and closeness centrality measures for threshold values 0.6, 0.7, and 0.8 are shown in Table 3.

\begin{table}[!htb]
\begin{center}
 \begin{tabular}{|c c c c|} 
 \hline
 Centrality/Threshold & 0.6 & 0.7 & 0.8 \\ [0.5ex]
 \hline
 Alpha & 96.37 & 93.8 & 93.17  \\ 
 \hline
 Eigenvector & 88.89 & 65.3 & 60.3 \\
 \hline
 Closeness & 76.69 & 61.51 & 42.32 \\
 \hline
\end{tabular}
\caption{Accuracy for alpha, eigenvector and closeness centrality measures for threshold values 0.6, 0.7 and 0.8}
\end{center}
\end{table}
For the simulated faults considered, it is also observed that the accuracy dropping with the threshold value increasing as the number of potential faulty nodes decreases. 

\noindent \textbf{\large{\emph{\\Research Questions}}} \\
In this paper, we address two questions that help in identifying faults in a distributed environment: \\
\newline
\normalsize{\textbf{Q1. }\emph{Does the choice of centrality index affect the accuracy of identifying the subset of potential faulty nodes for fault localization?\\}}
\newline
{As discussed, we have implemented three centrality-based indices used in fault localization, namely closeness, eigenvector, and alpha centrality measures. It is interesting to observe that alpha-centrality performs better for various threshold values than eigenvector and closeness centrality measures. Using alpha centrality, the calculation of node centrality values in weighted networks lets us consider the edge weights, a direct measure of the degree of influence of a fault on another fault. This results in better accuracy since the computation of centrality values is fault-specific and not cloud module-specific. It was alarming to observe that the closeness centrality measure produces the highest number of false-positive faults among the three considered centrality measures.} \\
\newline
\normalsize{\textbf{Q2. }\emph{In the graph model, how does the representation of faults as nodes improve the accuracy when compared to the representation of components as nodes?\\}}
\newline
{Even though having faults representing the vertices increases the complexity of the overall computation significantly due to one-to-many mapping between components and their faults, having a fault propagation-based graph reduces the number of false-positive cases as different faults in a component have different effects on other components. Faults representing nodes make the centrality calculation more specific to a fault, and hence we get a more accurate subset of potentially faulty nodes. This approach ensures both higher accuracy in choosing other modules when one faulty node is observed as well as prevents the selection of components with lesser probabilities of failing thus reducing the number of false-positive cases.}

\section{FUTURE WORK}

The cloud archetype has become more practical than ever and has been adopted by researchers, the IT industry, and other organizations. Fault-tolerance approaches are expected to improve services in the cloud environment. However, reliability issues of the cloud workflows have not been extensively considered in the existing approaches and evaluation has been done using some basic metrics, such as response time, availability, and throughput. Many studies model the system behaviour that includes execution paths and component interactions. A more complex, but interesting variant of this research work includes link failure between cloud components and fault tolerance of the same as well as employing more precise methods to compute $\alpha$ values in the centrality measure. Deep Learning approaches incorporated with centrality measures may be studied to predict fault-tolerance when the volume of the data increases. 

\section{CONCLUSION}
This research work has implemented centrality measure-based fault localization techniques to locate faults in distributed systems with a probabilistic method. In our experiments, we conducted tests using the suggested method with simulated faults in a distributed computing environment. The choice of centrality index impacts the precision of fault localization. The results of these experiments with eigenvector, closeness, and alpha centrality measures were analyzed. Among the three measures, it is observed that the identification of vulnerable cloud components has the highest accuracy with alpha-centrality measure and the least with closeness centrality measure and this sets a reliable way to prioritize vulnerable cloud endpoints and modules in an event of a fault occurring to make the architecture more robust.

\end{document}